\documentclass[global,twocolumn]{svjour}
\usepackage{graphicx}
\usepackage{dcolumn}
\usepackage{bm}
\usepackage{amsmath}

\journalname{Applied Physics B.}

\begin{document}

\title{Bandgap properties of two-dimensional low-index photonic crystals}

\author{Aaron Matthews\inst{1}\and Xue-Hua Wang\inst{1,2}\and Yuri Kivshar\inst{1}\and Min Gu\inst{3}}

\institute{Nonlinear Physics Center and Center for
Ultra-high bandwidth Devices for Optical Systems (CUDOS), Research
School of Physical Sciences and Engineering, Australian National
University, Canberra, ACT 0200, Australia,\\ 
phone: +61 2 6125 8277 fax: +61 2 6125 8588 \email{afm124@rsphysse.anu.edu.au}
\and Institute of Physics, Chinese Academy of Sciences, Beijing
100080, China \and Center for Micro-Photonics and Center for
Ultra-high bandwidth Devices for Optical Systems (CUDOS),
Swinburne University of Technology, P.O. Box 218, Hawthorn,
Victoria 3122, Australia}

\date{Received: date / Revised version: date}

\maketitle

\begin{abstract}
We study the bandgap properties of two-dimensional photonic
crystals created by a lattice of rods or holes conformed in a
symmetric or asymmetric triangular structure. Using the plane-wave
analysis, we calculate {\em a minimum value} of the refractive
index contrast for opening both partial and full two-dimensional
spectral gaps for both TM and TE polarized waves. We also analyze
the effect of ellipticity of rods and holes and their orientation
on the threshold value and the relative size of the bandgap.
\end{abstract}

\section{Introduction}

It is well established that three-dimensional periodic dielectric
structures, called {\em photonic crystals}, possess one or many
complete photonic band gaps in the transmission spectrum for
propagation of electromagnetic waves~\cite{book}.  Light-based
technological applications have their roots in planar technology,
which requires a knowledge of the bandgap properties of {\em
two-dimensional structures} assumed an infinitely long extension
in the third dimension. The theoretical study of photonic crystals
in two dimensions is easier due to the fact that in this case the
wave propagation can be analyzed separately for two different
polarizations, thus the original vector problem is reduced to two
scalar problems. These polarizations are Transverse Magnetic (TM),
if the electric field is perpendicular to the plane defining the
structure, and Transverse Electric (TE), if the same occurs for
the magnetic field. When the bandgaps for two different
polarizations overlap, they create a combined band gap known as an
absolute (or full) photonic bandgap. Several structures are known
to possess photonic band gaps for one of these polarizations and
for both polarizations simultaneously, as examined comprehensively
for high values of the refractive index ratio in the
two-dimensional geometry~\cite{wang4307JAP}.

One of the classical results in the theory of photonic crystals is
the existence of the critical (minimum) value of the refractive
index contrast to open a full spectral band gap in a
three-dimensional geometry~\cite{book}. For example, as was first
shown by Ho {\em et al.}~\cite{ho}, a diamond structure requires
the minimum refractive index contrast larger than 2. Similarly,
the threshold values of the refractive index appears in the theory
of two-dimensional photonic crystals~\cite{apl92}. In particular,
a triangular lattice of air holes in a dielectric material
possesses a large band gap for TE polarized waves and a complete
one for larger air holes.

Recently, several experimental groups demonstrated novel methods
for fabricating photonic crystals in solid polymer materials. In
particular, the group of Min Gu~\cite{ventura,straub} employed the
generation of submicron-size void channels by tightly focused
femtosecond-pulsed laser light; the technique is a one-step
approach which does not require chemical postprocessing, and it
allows to fabricate photonic crystals with a high degree of
perfection. In addition, the studies of Zhou {\em et
al.}~\cite{zhou} demonstrated two-dimensional triangular void
channel photonic crystals fabricated by femtosecond laser drilling
in a solid polymer material, and characterized their properties
for TE and TM polarized illumination. Although complete photonic
band gaps cannot exist in low-index contrast structures,
two-dimensional band gaps are possible for specifically polarized
electromagnetic modes, and such structures can be used for
photonic-crystal optical devices such as  superprisms or
waveguides.

The purpose of this paper is twofold. First, we calculate the
dependence of the spectral bandgaps on the refractive index
contrast for the two most popular types of triangular lattices of
two-dimensional photonic crystals. In particular, we find the
critical value of the refractive index contrast for opening
partial (for TM or TE polarized waves, respectively) as well as
full two-dimensional spectral band gaps. Second, being motivated
by the recent success in fabricating low-index photonic crystal
structures in solid polymers and chalcogenide glass, we explore
further the concept of the partial bandgaps of two-dimensional
photonic crystals and analyze the effect of ellipticity of rods
and holes and their orientation on the critical value and the size
of both partial and full spectral bandgaps.

As we assume that the materials we are working with are
macroscopic and isotropic, we are able to define the refractive
index as $n=\epsilon^{1/2}$ keeping $\mu =1$. In this paper, we
also interchange between the $\omega a/2\pi c$ form and the
$a/\lambda$ forms to show more clearly how the ratio increases
with a decreasing refractive index, as shown by Li {\em et
al.}~\cite{Li2574PRL}.

The paper is organized as follows. In Sec.~2 we consider a
two-dimensional photonic crystal created by a triangular lattice
of dielectric rods in air. In this case, a partial gap appears
first for the TM polarized waves, and it is shown to require a
relatively low index contrast. In Sec.~3 we consider the same
problem for a two-dimensional photonic crystal created by air
holes drilled in a dielectric slab, where the bandgaps first
appear for the TE polarized waves. And last, Sec.~4 concludes the
paper.

\section{Dielectric rods in air}

First, we consider a two-dimensional photonic crystal created by a
triangular lattice of circular or elliptic dielectric rods
assuming an arbitrary rotation of the elliptic rod relative to the
lattice symmetry axis. The photonic bandgap spectrum is calculated
by solving Maxwell equations by means of the plane-wave expansion
method~\cite{ho,plane_wave} employing the well-known numerical
algorithm~\cite{MIT_code}.

\begin{figure}
\includegraphics[width=80mm]{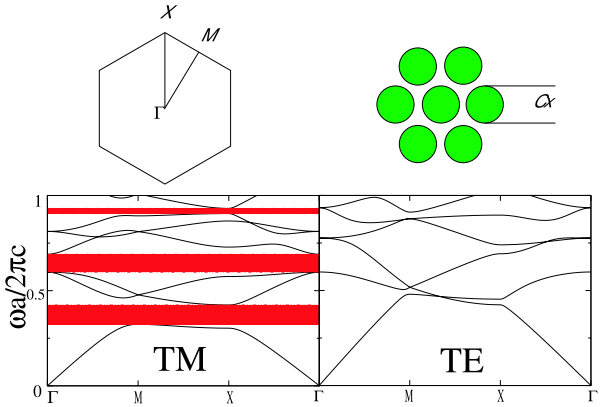}
\caption{Bandgap spectrum of TM (left) and TE (right) polarized
waves for a triangular lattice of circular rods ($C_x=C_y=0.575$)
at $\epsilon =5.8$ which is the permittivity of the chalcogenide
glass waveguides~\cite{ruan5140OE}.}
\label{fig1}
\end{figure}

\begin{figure}
\includegraphics[width=42.5mm]{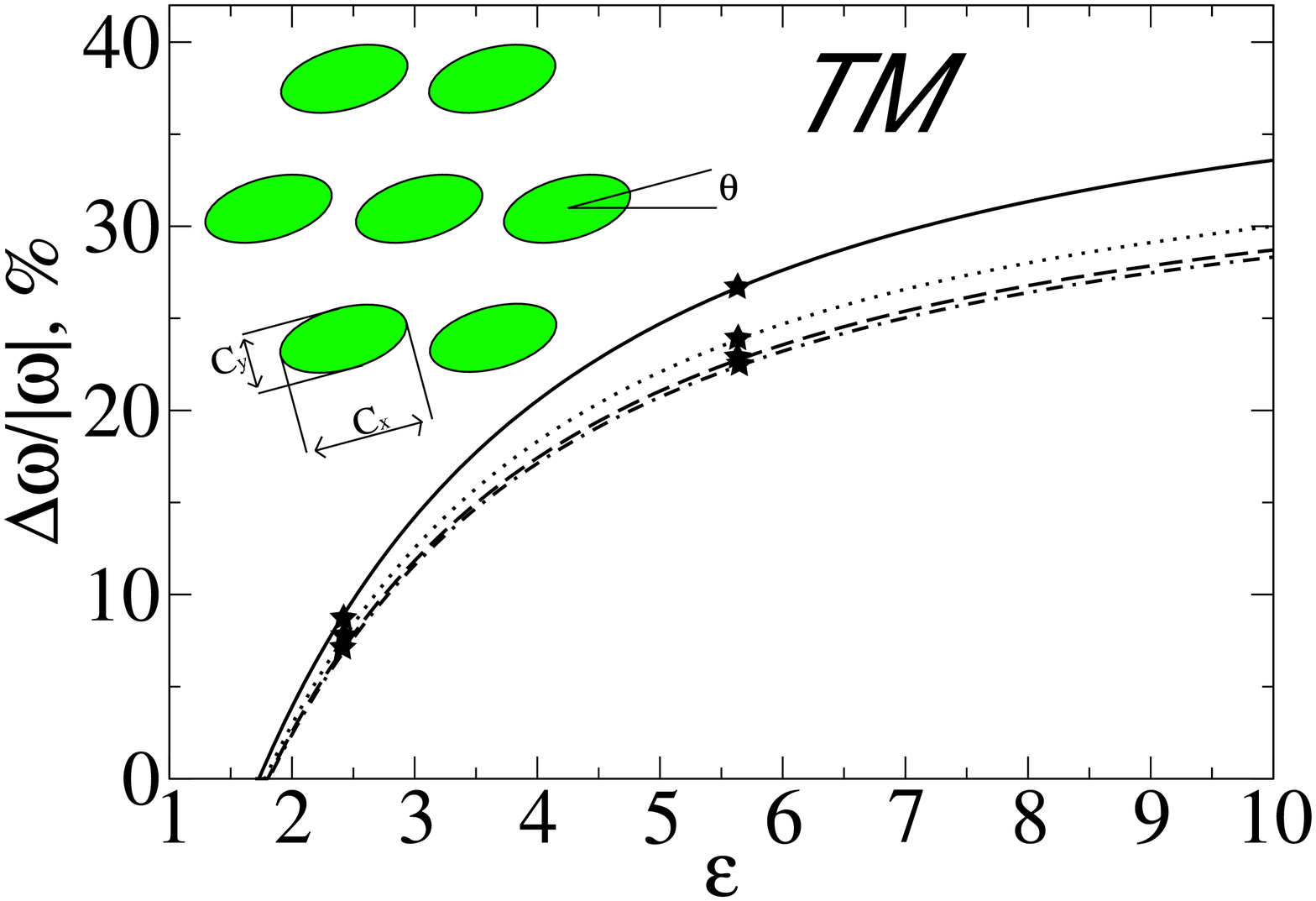}
\includegraphics[width=41.5mm]{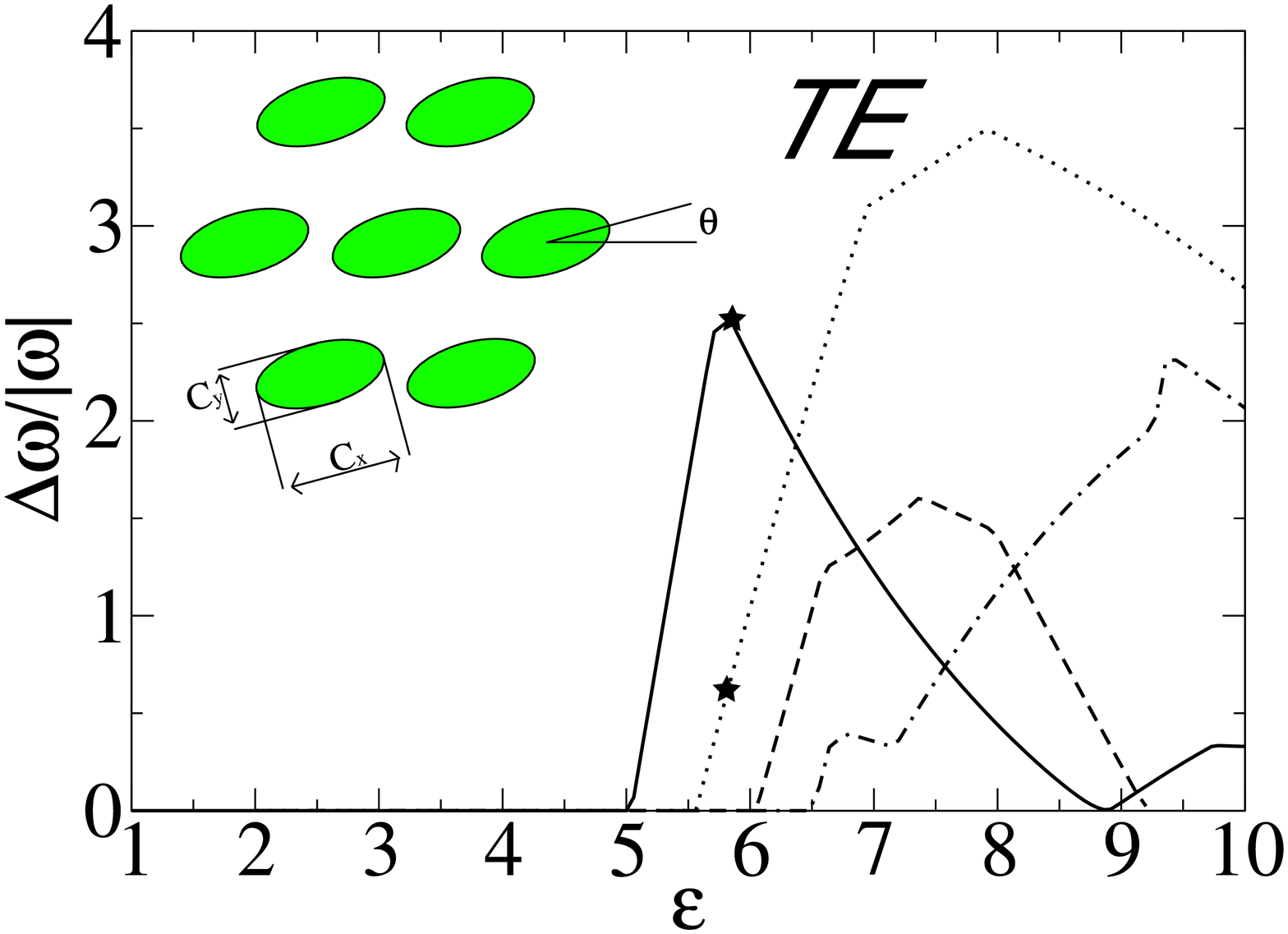}
\caption{(a,b) Relative size of the partial bandgaps for TM (left)
and TE (right) polarized waves for a triangular lattice of
circular (solid curve) and elliptic (other three curves) rods as a
function of $\epsilon$ for the filing factor 30\%. Dotted, dashed,
and dot-dashed curves show the results for the elliptic rods
($C_x=0.65$, $C_y=0.51$) with 0$^{o}$, 15$^{o}$, and 30$^{o}$
rotation angle. The critical value for the TM bandgap is $\epsilon
=1.73$. Stars indicate the permittivity of
polymer~\cite{straub} (at $\epsilon =2.4$) and the
permittivity of the chalcogenide glass
waveguides~\cite{ruan5140OE} (at $\epsilon =5.8$). Note the
significant difference in scale of the two graphs.} \label{fig2}
\end{figure}

An example of the photonic band-gap structure of such a
two-dimensional photonic crystal is shown in Fig.~\ref{fig1} for
the well-known case of a triangular lattice of circular rods. The
rods have the electric permittivity $\epsilon=5.8$ that
corresponds to the values measured for planar waveguides made of
chalcogenide glass~\cite{ruan5140OE}. In this case, the frequency
spectrum of a lattice of circular rods display several gaps for
the TM polarized waves, and two relatively large lower bandgaps
with the relative size 27.15\% and 14.61\%, respectively.

As the next step, we verify a general concept of the bandgap
spectrum of asymmetric lattices~\cite{our_laser} and consider a
triangular lattice made of elliptic rods with an arbitrary
orientation. In particular, we study the effect of the hole
rotation on the value of the partial and absolute bandgaps. These
results can naturally be compared with the bandgap spectra of the
two-dimensional structures created by circular holes (see below).
In a full agreement with the previous studies~\cite{pad} and
recent fabricated devices~\cite{fuji1478APL}, we observe that for
the case of dielectric rods in air a deviation of the cylinders
from a circular symmetry produces {\em a reduction} of the
relative size of the bandgaps. The similar effect is produced by
the rod rotation, so that larger values of the photonic band gap
are observed for the ellipses with smaller or no rotation (dotted
curve in Fig.~\ref{fig2}), and the bandgap becomes maximum for the
case of circular rods.

\begin{figure}
\includegraphics[width=80mm]{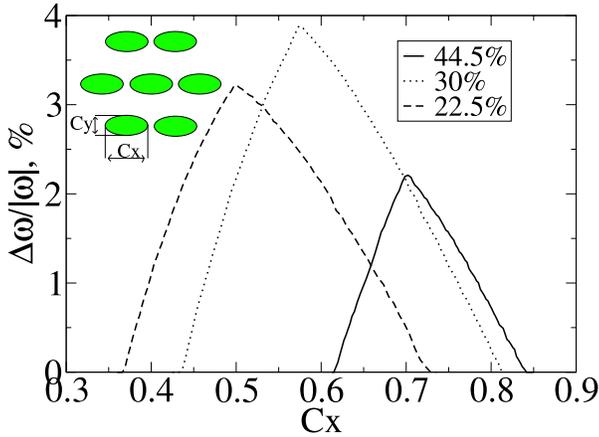}
\caption{Effect of varying rod ellipticity and filling fraction on
the size of the TM bandgap. Shown is the relative size of the
lower bandgap of the TM polarized waves (see Fig.~\ref{fig1}) as a
function of $C_x$, with $C_y$ varying to retain a constant filling
fraction, at $\epsilon=2$. Maximum bandgap is found for the
circles ($C_x=C_y=0.575$) with the filling fraction 30\%. }
\label{fig4}
\end{figure}

Finally, we study how the ellipticity of the dielectric rods in
the triangular-lattice photonic crystal may change the size of the
maximum TM bandgap at different values of the filling fraction. We
assume that the dielectric rods are ellipses with the axes $C_x$
and $C_y$, and we vary the value of $C_x$ for a fixed orientation,
also changing the size of $C_y$ in order to keep the filling
fraction constant. Figure~\ref{fig4} summarizes some of our
results for three values of the filling fraction, 44.5\% (solid),
30\% (dotted), and 22.5\% (dashed). The main result is that the
maximum value of bandgap is achieved for a triangular lattice of
circular rods (here, at $C_x=C_y=0.575$) with the filling fraction
30\%. This result is in agreement with all previous studies of
triangular-lattice two-dimensional photonic crystals.

\section{Air holes drilled in dielectric}

\begin{figure}
\includegraphics[width=80mm]{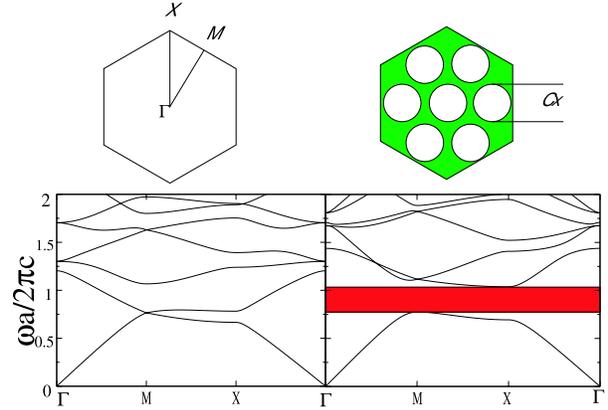}
\caption{Bandgap spectrum of TM (left) and TE (right) polarized
waves for a triangular lattice of circular holes ($C_x=C_y=0.735$)
at $\epsilon =5.8$ (the chalcogenide glass waveguides).}
\label{fig5}
\end{figure}

\begin{figure}
\includegraphics[width=41.5mm]{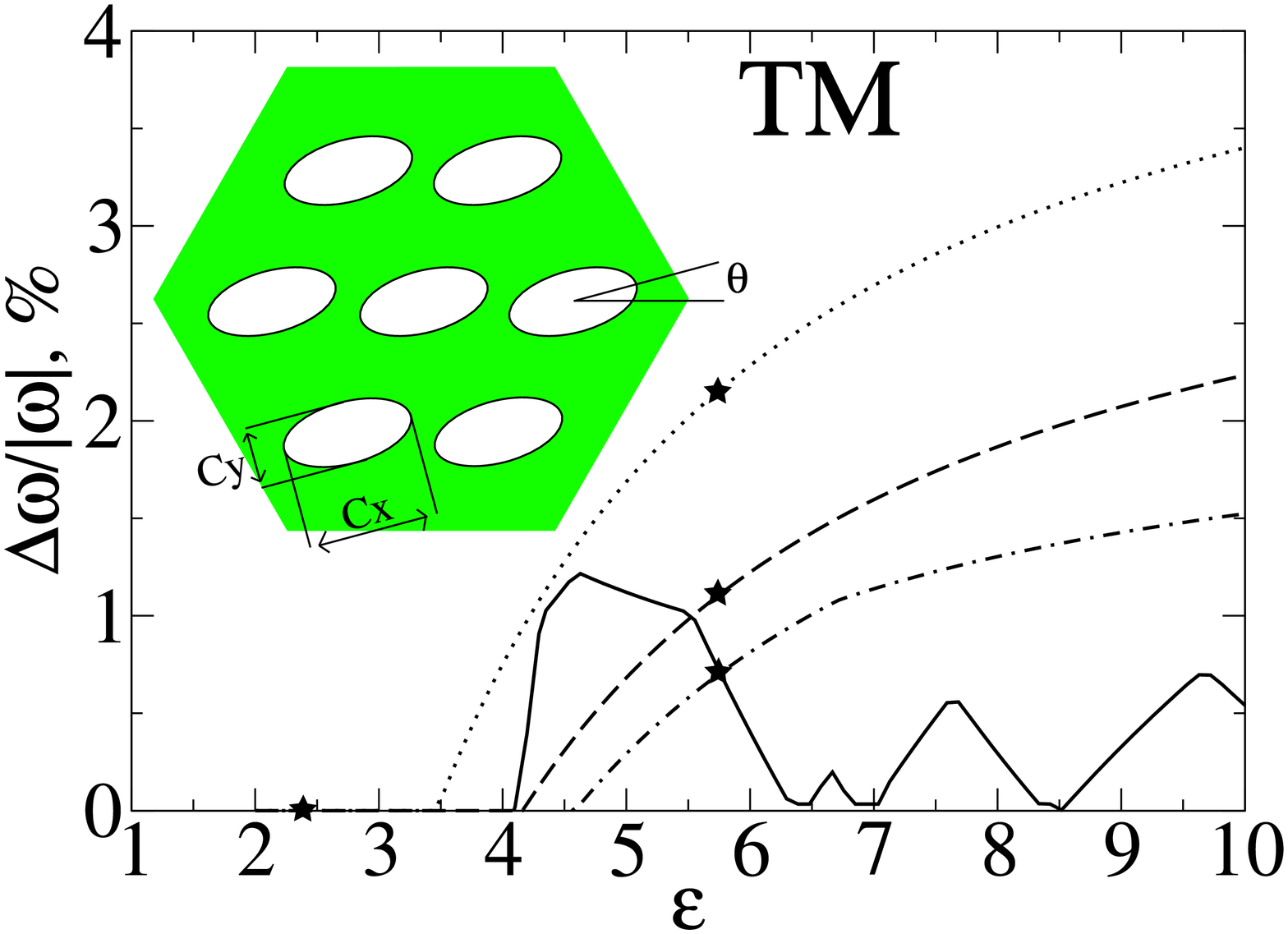}
\includegraphics[width=42.5mm]{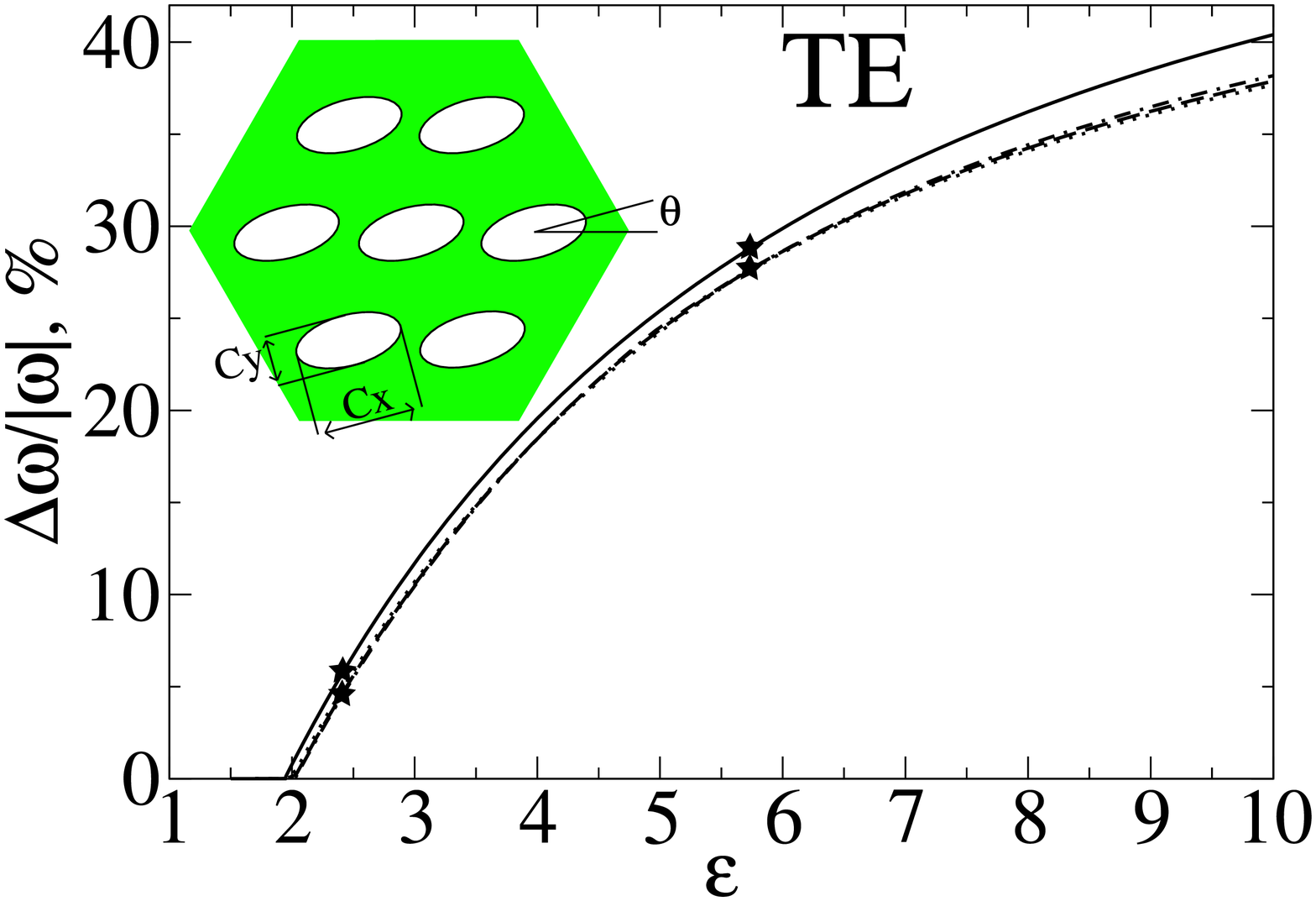}
\caption{(a,b) Relative size of the partial bandgaps for TM (left)
and TE (right) polarized waves for a triangular lattice of
circular (solid curve) and elliptic (other three curves) holes as
a function of $\epsilon$ and constant filing factor 51\%. Dotted,
dashed, and dot-dashed curves show the results for the elliptic
holes ($C_x=0.8$, $C_y=0.675$) with 0$^{o}$, 15$^{o}$, and
30$^{o}$ rotation angle. The critical value for the TE bandgap is
$\epsilon =1.95$. Stars mark the permittivity of the
polymer~\cite{straub} (at $\epsilon =2.4$) and the
permittivity of the chalcogenide glass
waveguides~\cite{ruan5140OE} (at $\epsilon =5.8$). Note the scale
difference of two graphs.} \label{fig6}
\end{figure}

Next, we consider the other important case when a two-dimensional
photonic crystal is created by a triangular lattice of circular
[see Fig.~\ref{fig5}] or elliptic [see Fig.~\ref{fig6}] holes
drilled in a dielectric slab, assuming an arbitrary rotation of
the elliptic hole relative to the lattice symmetry axis.

As opposed to the case of rods, air holes produce an extremely
large bandgap in the TE spectrum of a photonic crystal, this is
demonstrated experimentally in the work of zhou {\em et al.}~\cite{zhou}. The key
advantage to this system as opposed to rods is the frequency
position of the bandgap, especially for the case shown in
Fig.~\ref{fig5}, which is the case of the refractive index
corresponding to the chalcogenide glass
waveguides~\cite{ruan5140OE}. When it is viewed as a wavelength
using the simple conversion $\omega a/2\pi c = a/\lambda$ we can
see that with the bandgap around $a/\lambda =1$, the wavelength of
the confined light is equal to the size of the lattice giving us
the ability to fabricate the planar structures on the scale of the
wavelength of interest unlike the common $a/\lambda$ values of 0.3
to 0.5 which would lead to structures 1/3 to 1/2 of the lattice
size in order to produce a bandgap for the same wavelength, a
challenge for any fabrication method.

Another advantage is in the nature of the holes drilled in a
dielectric structure such as physical resistance to a damage.
Indeed, the rods which, when being fabricated with the large
aspect ratios to exhibit bandgaps, are very sensitive to a
physical damage, whereas the hole structure, which has a constant
lattice of joined dielectric is far stronger and therefore more
resistant to the rigors of the fabrication processes than
free-standing rods.

For the photonic crystals fabricated in the chalcogenide glass
waveguides~\cite{ruan5140OE}, round holes, 1139nm in diameter,
forming a trigonal lattice with a lattice spacing of 1550nm should
provide a photonic bandgap at telecommunications wavelengths. This
should also, as shown by the previous studies, be able to support
different guided modes although the adaptation of the structure to
the specific optical devices should be analyzed in more details.

As follows from Figs.~\ref{fig6}(a,b), the tradeoff when
optimizing the structures also applies to the hole photonic
crystals. In this case, the effect of rotation is insignificant
and the case of purely circular holes is optimal and produces a
bandgap of over 40\% at $\epsilon =10$. When comparing
Fig.~\ref{fig2} and Fig.~\ref{fig6}, we can see that the TM
bandgap for the rod structure starts at a lower value of the
permittivity than the TE bandgap for holes, and it can be seen
that the rod bandgap is larger than the hole bandgap until around
$\epsilon =2.8$ after which the hole bandgap becomes significantly
larger $(\approx 7\%$ at $\epsilon = 10)$. This should be taken
into consideration when choosing an appropriate structure for a
given fabrication process.

\begin{figure}
\includegraphics[width=80mm]{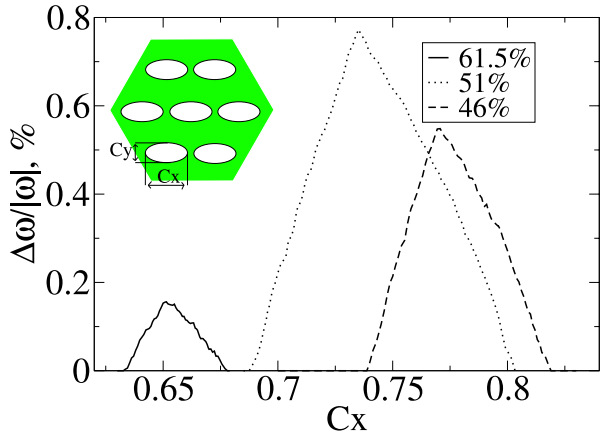}
\caption{Effect of varying hole ellipticity and filling fraction
on the size of the TE bandgap. Shown is the relative size of the
lower bandgap of the TE polarized waves (see Fig.~\ref{fig5}) as a
function of $C_x$, with $C_y$ varying to retain constant filling
fraction at $\epsilon=2$. Maximum band gap is found for the
circles ($C_x=C_y=0.735$) with the filling fraction 51\%. }
\label{fig8}
\end{figure}

Figure~\ref{fig8} shows, by a very steep drop-off on all curves,
that ellipticity has a far greater effect in this system than on
the system shown in Fig.~\ref{fig4}. It should be remembered in
these two graphs that after the system passes through the point
corresponding to a circle it becomes an equivalent ellipse rotated
by $30^o$. While this will prove a problem for fabrication the
bandgaps are large enough that small errors in ellipticity will
not lead to the removal of the bandgap completely. It can also be
seen that an error towards larger holes, visualized as a lower
filling fraction of dielectric, retains the bandgap size better
than a reduction in the hole size.

Our results demonstrate that a threshold value of the permittivity
to open a partial bandgap is 1.73 which equates to a refractive
index of 1.31, and we therefore believe that it is safe to rule
out attempts at producing photonic bandgap devices in any system
with a refractive index lower than this limit.

While ellipticity caused by directional defocusing of the etching
beam causes rapid loss of bandgap, implementing the optimal system
allows a much greater resistance to ellipticity using the simple
fact that the change in bandgap size vs. ellipticity is fairly
constant as we move further from the maximum so an advantageous
filling factor will lead to a larger margin for error before the
bandgap disappears.

\section{Conclusions}

We have analyzed the bandgap properties of two-dimensional
photonic crystals created by triangular lattices of dielectric
rods in air and air holes drilled in a planar dielectric slab, for
different (including the lowest possible) values of the refractive
index contrast. Using the plane-wave analysis, we have calculated
{\em the critical value} of the refractive index contrast for
opening partial (either for TM or TE polarized waves) and full
two-dimensional spectral bandgaps. We have analyzed the effect of
ellipticity of rods and holes and their orientation on the
critical value and the size of the bandgaps. In particular, we
have predicted that partial bandgaps  may appear in the frequency
spectrum for the index contrast as low as $\epsilon= 1.73$, in the
case of rods (for the TM polarized waves), and $\epsilon= 2$, in
the case of holes (for TE polarized waves). We have demonstrated
also that, by reducing the refractive index from some large values
(e.g. for Si) to lower value slightly above the threshold, we are
able to obtain far more fabricable periodic structures for
experiment due to an increase in the wavelength-to-period ratio.
We believe that our results will be important for the current
efforts in fabricating planar photonic-crystal structures based on
dielectric materials with low refractive index such as solid
polymers, polymer resin, and chalcogenide glasses.

\section*{Acknowledgements}

The authors thank Sergei Mingaleev for useful discussions and
references. This work was supported by the Australian Research
Council through the Center of Excellence Program.

\end{document}